\documentclass[review]{elsarticle}
\usepackage{lineno,hyperref}
\usepackage{subfigure}
\usepackage{epsfig,graphics,subfigure,psfrag,amsmath,amssymb}
\usepackage{amsmath,amssymb,amsfonts}
\usepackage{algorithmic}
\usepackage{graphicx}
\usepackage{textcomp}
\usepackage[vlined,boxed,ruled]{algorithm2e}
\usepackage{dsfont}
\newtheorem{myDef}{Definition}

\begin{document}

\begin{frontmatter}

\title{A Privacy-Preserving Traffic Monitoring Scheme via Vehicular Crowdsourcing}

\author[a]{Chuan Zhang}
\author[a]{Liehuang Zhu\corref{mycorrespondingauthor}}

\cortext[mycorrespondingauthor]{Corresponding author}
\ead{liehuangz@bit.edu.cn}

\author[a]{Chang Xu\corref{mycorrespondingauthor}}

\ead{xuchang@bit.edu.cn}

\author[b]{Xiaojiang Du}

\author[c]{Mohsen Guizani}

\address[a]{Beijing Engineering Research Center of Massive Language Information Processing and Cloud Computing Application, School of Computer Science and Technology, Beijing Institute of Technology, Beijing, China.}

\address[b]{Department of Computer and Information Sciences, Temple University, Philadelphia, USA.}

\address[c]{College of Engineering, Qatar University.}

\begin{abstract}
The explosive growth of vehicle amount has given rise to a series of traffic problems, such as traffic congestion, road safety, and fuel waste. Collecting vehicles' speed information is an effective way to monitor the traffic condition and avoid vehicles being congested, which however may bring threats to vehicles' location and trajectory privacy. Motivated by the fact that traffic monitoring does not need to know each individual vehicle's speed and the average speed would be sufficient, we propose a privacy-preserving traffic monitoring (PPTM) scheme to aggregate vehicles' speeds at different locations. In PPTM, the roadside unit (RSU) collects vehicles' speed information at multiple road segments, and further cooperates with a service provider to calculate the average speed information for every road segment. To preserve vehicles' privacy, both homomorphic Paillier cryptosystem and super-increasing sequence are adopted. A comprehensive security analysis indicates that the proposed PPTM can preserve vehicles' identities, speeds, locations, and trajectories privacy from being disclosed. In addition, extensive simulations are conducted to validate the effectiveness and efficiency of the proposed PPTM scheme.
\end{abstract}

\begin{keyword}
\texttt{traffic monitoring; speed; privacy-preserving; vehicular crowdsourcing}
\end{keyword}

\end{frontmatter}

\section{Introduction}
Nowadays, the number of global vehicles has exceeded 1.2 billion and may be headed to 2 billion by 2035 \cite{adcs}. With such a massive amount of vehicles, many critical social problems, such as traffic congestions and slow traffic, have emerged, leading to significant time and fuel waste. According to a report released by Harvard Center, for the drivers in 10 most-congested cities in USA, more than 48 hours are wasted in traffic jams, causing \$121 billion loss in time and fuel every year \cite{trafficjam}. To deal with these critical problems, both industry and academia are paying great attention to traffic monitoring, and vehicular ad hoc network (VANET) is considered as one of the most promising way that can be leveraged in traffic management \cite{WuJLOK15, AbboudZ16}.

In VANETs, vehicles, embedded with onboard units (OBUs), can share traffic information (e.g., locations and speeds) to the roadside units (RSUs) through vehicle-to-infrastructure (V2I) communications, and nearby vehicles by vehicle-to-vehicle  (V2V) communications \cite{zhuzxdxsm}.  By collecting and analyzing these traffic information, vehicles can easily know different locations' traffic conditions and road safety, and accordingly plot out their optimal routes. Recently, several VANET-based traffic monitoring applications have been built. For example, Google and Apple provide real-time navigation services based on current traffic information \cite{WuZPM16}. Waze has developed an application that can help drivers get the best route with real-time help from other drivers \cite{waze}. Although many benefits can be brought by this emerging network paradigm, its flourish still hinges on how to resolve security and privacy concerns for the drivers. Since a vehicle's location is tightly bundled with its driver, an attacker can predict a driver's future location based on his vehicle's trajectory, or even infer the driver's personal information, such as habits, health condition, wage income, religious belief, according to its frequent visiting places.

\begin{figure}
	\centering
	\includegraphics[ width=12cm]{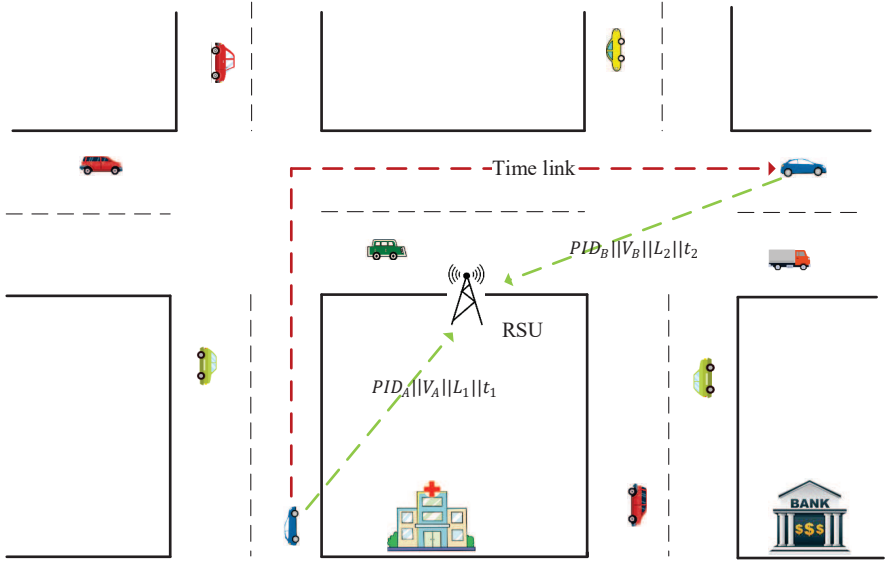}
	\caption{Using passing time to link pseudonyms}
	\label{fig:linkable}
\end{figure}

To preserve the vehicles' privacy, pseudonyms and anonymous authentication are two effective ways to conceal vehicles' real identities and realize conditional privacy preservation \cite{XuXYH18,YangXXWLY19,LwamoZXSLZ19,DuC08}. For example, Ni et al. \cite{NiLZS16} proposed a privacy-preserving real-time navigation system by collecting vehicles' location and speed information, and with the randomization technique, the sensitive identity privacy is preserved. However, the work in \cite{GolleP09} showed that user identities can sometimes be inferred from the location data if users' home and work locations are deduced from the data. Moreover, we observe an attack that by linking vehicles' speed information, vehicles can also be identified even if they change their pseudonymous. An example is illustrated in Fig. \ref{fig:linkable}. At time $t_1$, a vehicle provides its speed information $PID_A||v_1||L_1||t_1$ to a roadside unit (RSU), and at time $t_2$, it uploads another speed information $PID_B||v_2||L_2||t_2$, where $v$ denotes the average speed in the road segment, $L$ denotes the location, and $t$ represents the current time. Although the vehicle's pseudonym is changed (i.e., $PID_A \rightarrow PID_B$), attackers can still link the pseudonyms by comparing the estimated passing time \footnote{The average passing time can be calculated by using the distance and average speed. The distance from $L_1$ to $L_2$ can be obtained from GPS.} and actual passing time (i.e., $t_2 - t_1$) between these two locations. Thus, there still lacks a privacy-preserving traffic monitoring scheme which can protect the vehicles' identities and defend against the linkable attack.

In this paper, to deal with the above challenges, we propose a privacy-preserving traffic monitoring (PPTM) scheme to enable vehicles provide their traffic information while not sacrificing their privacy. This scheme uses the homomorphic Paillier cryptosystem to guarantee the privacy of vehicles' speeds, and adopts a well defined super-increasing sequence to not only protect vehicles' location privacy, but also save tremendous computational costs and communication overhead. Our main contributions can be further summarized below.
\begin{itemize}
	\item First, inspired by the fact that traffic monitoring does not need to know each individual vehicle's speed and the average speed would be sufficient, we propose PPTM which uses the super-increasing sequence and homomorphic Paillier cryptosystem to realize privacy-preserving speed aggregation and efficient traffic monitoring. Concretely, each vehicle uses a well defined super-increasing sequence to aggregate its multiple speeds
	and encrypts the aggregated result before uploading to the RSU. Then, the RSU will aggregate all reports and cooperate with a service provider to calculate different road segment's average speed. During this process, vehicles' identity, speed, and location privacy will not be disclosed to any other party.
	\item Second, we find that the anonymous technologies such as pseudonyms and randomizable signature are not suitable for certain VANET-based applications because of the time link attack. To mitigate this attack, we design a privacy-preserving data aggregation approach. Through a comprehensive security analysis, the proposed PPTM is proved to be secure and privacy preservation. Particularly, the proposed scheme can achieve report privacy preservation, report authentication and data integrity, identity preservation, and can also defend against the collusion attack. The detailed analysis is given in Section 5.
	
	\item Third, we conduct extensive simulations to show PPTM is practical and efficient. Compared with a traditional baseline scheme, PPTM can significantly reduce computational costs and communication overhead, indicating that the proposed scheme can indeed realize real-time traffic monitoring.
\end{itemize}

The rest of this paper is organized as follows. In section 2, we introduce the system model, security requirement, and design goals of the proposed PPTM scheme. In section 3, preliminaries including bilinear pairings and Paillier cryptosystem are introduced. The detailed introduction of PPTM is given in section 4. In section 5 and section 6, we analyze the security and performance of PPTM respectively. In section 7, some related works are listed, and we draw our conclusion in section 8.
\section{System Model, Security Requirements, and Design goals}
In this section, we formalize the proposed scheme by giving the system model, threat model, and design goals.

\subsection{System Model}
\begin{figure}
	\centering
	\includegraphics[width=10cm]{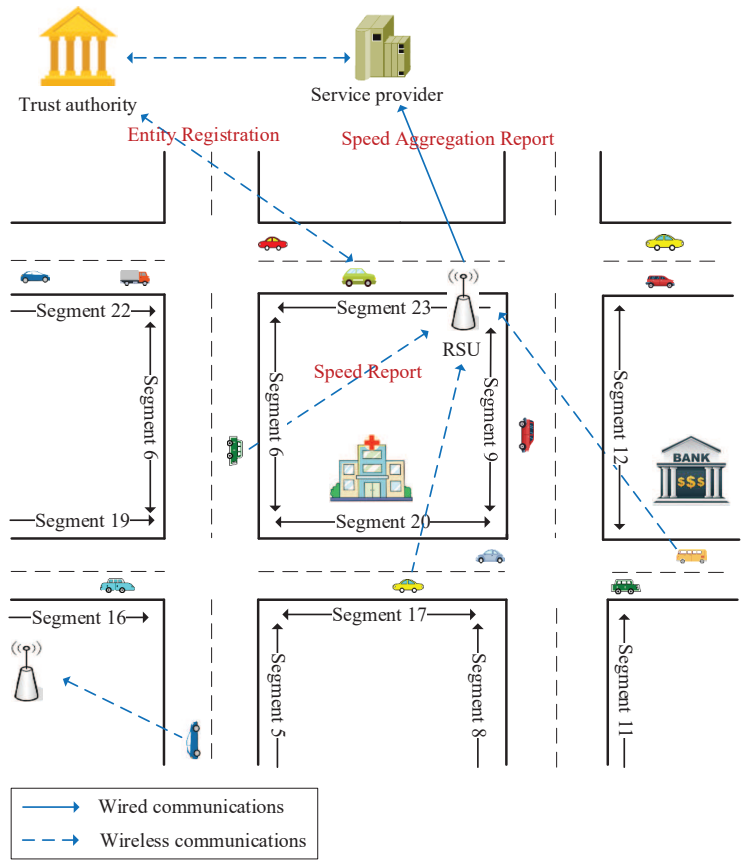}
	\caption{Using trust values to link pseudonyms in a given period of time.}
	\label{fig:system}
\end{figure}

In the proposed PPTM scheme, roads are divided into multiple segments and vehicles are expected to provide their average speed for each segment they have passed through. A typical RSU-assisted VANET application is illustrated in Fig. \ref{fig:system}. In particular, the considered system model consists of the following entities.

\begin{itemize}
	\item Trust authority (TA): TA is a fully trusted entity which is responsible for the registration of vehicles and RSUs. It builds public/secret key pairs for all entities, and generates sufficient pseudonyms for vehicles before their next registration.
	\item Vehicles: All vehicles are embedded with onboard units (OBUs) which enable them communicate with RSUs and surrounding vehicles through wireless communications. Besides, they also have the ability to generate and run their own homomorphic cryptosystem.
	\item Roadside units (RSUs): RSUs act as the role of access points which are widely deployed in the urban area. They can communicate with vehicles via wireless communications and the service provider by wired communications.
	\item Service provider (SP): SP is a centralized entity which is responsible to provide traffic monitoring services. It connects with all RSUs through fast communication technology, such as wired cables.
\end{itemize}

\subsection{Security Requirements}
In our security model, TA is fully trusted as it is responsible to initialize the whole system and generate credentials and public/private keys for all participating entities. SP and RSUs are considered to be honest-but-curious, which means both of them will strictly follow the designed protocol, but are curious about vehicles' privacy. In particular, we assume there is no collusion between SP and RSUs, which is similar to most existing RSU-assisted scenarios \cite{ZhangZXSDG19,XueHMWHY18}. Meanwhile, we assume that vehicles will provide correct speed information to the RSU. This assumption is reasonable in most traffic monitoring scenarios, since 1) the speed provided by vehicles is in the area where they have passed through, and providing false data would not benefit them, and 2) vehicles want to know the correct traffic conditions, and thus will honestly follow the designed protocol for their mutual benefits. Besides, we also assume there exists an attacker which is curious about drivers' privacy. It may launch attacks, modify speed reports, and threat data integrity. Based on the above assumptions, the proposed scheme should achieve the following security requirements.

{\em Identity Privacy Preservation.} As described earlier, an attacker can potentially identify drivers even though they adopt pseudonyms and anonymous authentication. Thus, to preserve drivers' identity privacy, attackers cannot infer vehicles' location  information (i.e., road segments) based on the given data.

{\em Location Privacy Preservation.} Since the speed is location-aware, preserving drivers' location privacy requires preventing their speed from being disclosed. Hence, the proposed scheme should ensure that even if the RSU or an attacker receives a vehicle's speed information, it cannot recover its speed and further infer its location privacy.

{\em Data Integrity.} An attacker may eavesdrop drivers' reports and modify them for its benefits. Thus, the proposed scheme should guarantee data integrity and any malicious operations should be detected.

\subsection{Design goals}
Based on the aforementioned security requirements, our goal is to design a privacy-preserving traffic monitoring scheme, which enables vehicles upload their speeds towards the RSU securely and efficiently. Concretely, the proposed scheme should achieve the following two design goals.

{\em The defined security requirements should be guaranteed.} If the proposed scheme fails to realize the aforementioned security requirements, drivers' identity and location privacy may be disclosed, and data reports transmitted to the RSU or other vehicles may be modified. Then, vehicles may be reluctant to provide their speed, and traffic conditions will not be accurately estimated.

{\em High efficiency should be guaranteed.} To provide real-time traffic monitoring, vehicles are expected to upload speed information in a short transmission interval. However, to preserve drivers' privacy, the sensitive information should be encrypted, which may introduce tremendous computational costs and bandwidth consumption for the resource-constrained vehicles. Thus, the proposed scheme should achieve high efficiency in computational costs and communication overhead.

\section{preliminaries}
In this section, we review the pairing-based cryptography \cite{BonehF03} and the Paillier cryptosystem \cite{Paillier99}, which serve as the basis of our proposed traffic monitoring scheme.

\subsection{Bilinear Pairings}
Suppose there are two cyclic groups $\mathbb{G}_1$ and $\mathbb{G}_2$, both of which share a same order $q$. Then, a bilinear map $e: \mathbb{G}_1 \times \mathbb{G}_1 \rightarrow \mathbb{G}_2$ has the following properties.
\begin{itemize}
	\item Bilinearity: $e(aP,bQ) = e(P,Q)^{ab} \in \mathbb{G}_2$, for all $P,Q \in G_1$ and $a,b \in \mathbb{Z}^*_{q}$.
	\item Non-degeneracy: $e(P,P) \neq 1$, for all $P \in \mathbb{G}_1$.
	\item Computability: $e(P,Q)$ can be efficiently computed, for all $P,Q \in \mathbb{G}_1$.
\end{itemize}

By referring to \cite{AbdallaBR01,XuLWZH17}, we give two more comprehensive definitions for bilinear pairings.

\begin{myDef}
	Given an input security parameter $\kappa$, $\mathcal{G}en$ is a probabilistic algorithm to output a 5-tuple $(q, P, \mathbb{G}_1, \mathbb{G}_2, e)$, in which $q$ is a $\kappa$-bit prime, $P$ is a generator, $(\mathbb{G}_1, \mathbb{G}_2)$ are two cyclic groups sharing a same order $q$, and $e: \mathbb{G}_1 \times \mathbb{G}_1 \rightarrow \mathbb{G}_2$ is an efficient, computable, and non-degenerated bilinear map.
\end{myDef}

\begin{myDef}[Computational Diffie-Hellman (CDH) Problem]
	Given elements $(P,aP,bP) \in \mathbb{G}_1$, there exists no effective algorithm can calculate $abP \in \mathbb{G}_1$ for unknown $a,b \in \mathbb{Z}^*_q$ in a probabilistic and polynomial time.
\end{myDef}

\subsection{Paillier cryptosystem}
As an effective technology to achieve homomorphic properties on the ciphertexts, Paillier cryptosystem has been widely used in various privacy-preserving applications. Concretely, three algorithms are included in the Paillier cryptosystem.
\begin{itemize}
	\item {\em Key Generation:} With a security parameter $\kappa_1$, select two large $\kappa_1$-bit primes $p_1, q_1$, and calculate $n = p_1 q_1$ and the least common multiple of $p_1$ and $q_1$, i.e., $\lambda = lcm(p_1,q_1)$. Then, define a function $L(a) = \frac{a-1}{n}$, and calculate $\mu = (L(g^\lambda \ \text{mod} \ n^2))^{-1} \ \text{mod} \ n^2$, where $g \in \mathbb{Z}^*_n$. Then, the public/private keys are $pk=(n,g)$ and $sk=(\lambda,mu)$.
	\item {\em Message Encryption:} Given a plaintext $m \in \mathbb{Z}_n$, after choosing a random value $r \in \mathbb{Z}^*_n$, the message is encrypted as $c = E(m) = g^m \cdot r^n \ \text{mod} \ n^2$.
	\item {\em Ciphertext Decryption:} Given a ciphertext $c = E(m) \in \mathbb{Z}^*_{n^2}$, the message is recovered as $m = D(c) = L(c^{\lambda \ \text{mod} \ n^2}) \cdot \mu \ \text{mod} \ n$.
\end{itemize}

Note that, Paillier cryptosystem has been proven to be correct, secure, and effective against the chosen plaintext attack \cite{Paillier99}. Moreover, Paillier cryptosystem allows arithmetic operations on ciphertexts, such as $E(m_1)\cdot E(m_2) = E(m_1 + m_2)$ and $E(m_1)^a = E(a \cdot m_1)$, for all $(m_1, m_2) \in \mathbb{Z}^*_n$.

\section{Proposed PPTM Scheme}
In this section, we will give the details of the proposed PPTM scheme which includes system initialization, speed request and speed reporting, privacy-preserving report aggregation, secure report reading, and traffic guidance and identity tracing.
\subsection{System Initialization}
TA initializes the whole system. After selecting two security numbers $\kappa, \kappa_1$, it first runs $\mathcal{G}en(\kappa)$ to generate a 5-tuple  $(q, P, \mathbb{G}_1, \mathbb{G}_2, e)$ and calculates public/private keys of the Paillier cryptosystem, i.e.,  $pk = (n,g), sk = (\lambda, \mu)$, according to $\kappa_1$. Then, TA selects a secure cryptographic hash function $H$, where $H: \{0,1\}^* \rightarrow \mathbb{G}_1$. Vehicles are required to register themselves at set intervals. TA chooses a secure key $k_0$ and generates a secure symmetric encryption algorithm $AES_{k_0}$. For every registered vehicle with its real identity number $ID_i$ \footnote{The real ID can be license number or social secure number.}, TA generates a group of pseudonyms $\{PID_{ij} = AES_{k_0}(ID_i || x_{ij})\}^{n}_{j=1}$ by choosing  a set of random values $\{x_{ij}\}^n_{j = 1 } \in \mathbb{Z}^*_{q}$. Then, TA uses $x_{ij}$ as each vehicle's certified public key and calculates the corresponding private key as $Y_{ij} = x_{ij}P$. For each RSU with its identity number $ID_r$, TA selects a random number $x_r \in \mathbb{Z}^*_q$ as its public key and calculates the private key as $Y_r = x_r P$. Hence, TA sends $\{\{PID_{ij}, x_{ij},Y_{ij},\}^n_{j=1}, (P, \mathbb{G}_1, \mathbb{G}_2, e, H), (n,g), (IDr,Y_r)\}$ to each vehicle, $\{(n,g), ID_r,x_r, Y_r\}$ to each RSU, and $(\lambda, \mu, (ID_r,Y_r))$ to SP.

In addition, in the coverage of a RSU, roads are divided into multiple segments. Assume that the maximum number of segments within the coverage of each RSU is $M$, the number of vehicles in every segment is no more than $Q$, and the maximum speed in every road segment is smaller than $V$. Then, for the segments located in each RSU's coverage, TA generates a super-increasing sequence $\overrightarrow{a} = (a_1,a_2, \cdots, a_M)$, where $a_i$ denotes the $i$-th segment such that $a_1 \in \mathbb{Z}^*_n$ is randomly chosen, $\sum^{j-1}_{i=1} a_i \cdot Q \cdot V < a_j$ for $j = 2, 3. \cdots, M$, and $\sum^{M}_{i=1} a_i \cdot Q \cdot V < n$.

\subsection{Speed Request and Speed Reporting}
\begin{figure}
	\centering
	\includegraphics[width=10cm]{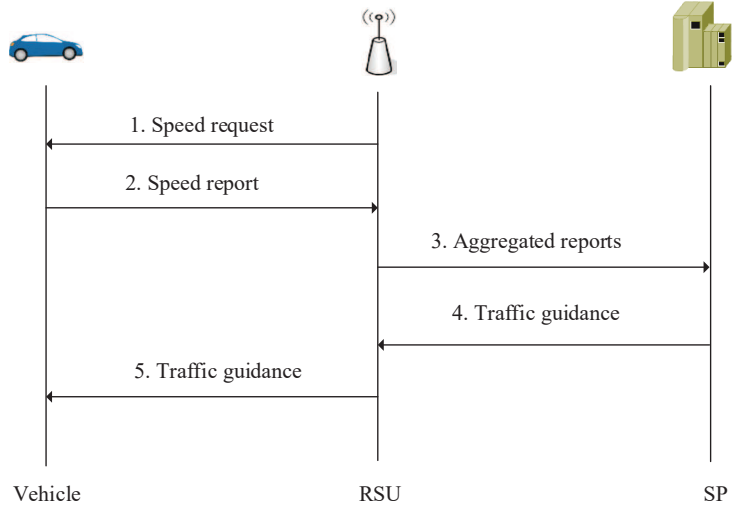}
	\caption{System procedure of PPTM}
	\label{fig:procedue}
\end{figure}

Fig. \ref{fig:procedue} illustrates the system procedure of PPTM. As can be seen, RSU first generates a speed request and all vehicles response it by providing their driving reports. Specifically, the request contains the RSU's id, the current timestamp $TS$, time range $TR$, and the signature $\sigma_r = x_r H(ID_r||TS||TR)$. Note that, the timestamp is used to defend against the replay attack launched by other forged RSUs. Then, the RSU broadcasts the request $R_r = ID_r||TS||TR||\sigma_r$ to vehicles driving in its communication coverage. After receiving this request, vehicles first verify the report by examining whether $e(P,\sigma_r)$ equals to $e(Y_r, H(ID_r||TS||TR))$. If the equation holds, the request will be accepted, since $e(P,\sigma_r) = e(x_r P, H(ID_r||TS||TR)) = e(Y_r, H(ID_r||TS||TR||))$.

If the request is valid, vehicles are expected to provide their speed reports. The format of speed is defined as $\{(A_i(j), S_i(j))\}^{M,V}_{i=1,j=1}$, where $t_i(j)$ denotes the time passing by the $i$-th segment for the vehicle $\mathcal{V}_j$, and $\{A_i(j), S_i(j)\}$ are calculated as follows,
\small
\begin{equation}
	\notag
	\begin{aligned}
		\left \{ \begin{aligned}
			A_i(j) & = \left \{ \begin{array} {lr}
				1 \ \ \text{if} \ \mathcal{V}_j \ \text{has passed by the segment} \ i\\
				0 \ \ \text{if} \ \mathcal{V}_j \ \text{has not passed by the segment} \ i  \\
			\end{array}
			\right.\\
			S_i(j) &= \left \{ \begin{array}{lr}
				v_i(j) \ \ \text{if} \ \mathcal{V}_j \ \text{has passed by the segment} \ i\\
				0 \ \ \text{if} \ \mathcal{V}_j \ \text{has not passed by the segment} \ i  \\
			\end{array}
			\right. \\
		\end{aligned}
		\right.
	\end{aligned}
\end{equation}
\small
where $v_i(j)$ is $\mathcal{V}_j$'s average speed when passing by the segment $i$. For example, if a vehicle's speed report is expressed as $\{(5,1,50),(0,0,0),(0,0,0),(0,0,0),(3.5,1,75),(2,1,60)\}$, this means that during the past 10.5 minutes the vehicle has passed by the segment 1, 5, and 6, with the average speed as 50, 75, and 60. Then, given a time range as 8, the vehicle should submit the speed report $\{(3.5,1,75),(2,1,60)\}$ as $3+2.5<8$. To preserve the privacy of location and speed privacy, the report should be encrypted before uploading to the RSU. The vehicle $\mathcal{V}_j$ selects two random values $r_{j1}, r_{j2} \in \mathbb{Z}^*_{n}$ and calculates the ciphertexts as $C_{j1} = g^{(a_1 \cdot A_1(j) + \cdots + a_M \cdot A_M(j))} \cdot r^n_{j1}\ \text{mod} \ n^2$ and $C_{j2} = g^{(a_1 \cdot S_1(j) + \cdots + a_M \cdot S_M(j))} \cdot r^n_{j2} \ \text{mod} \ n^2$. Then, the vehicle signs the report with its secret key and timestamp by computing $\sigma_j = x_j H(PID_j || Y_j || C_{j1}||C_{j2}||TS)$. After that, $\mathcal{V}_j$ delivers the speed report $R_j = PID_j|| Y_j||C_{j1}||C_{j2}||TS||\sigma_j$ to the RSU.

\subsection{Privacy-Preserving Report Aggregation}
Upon receiving the report, RSU first checks the freshness of this report, i.e., to make sure that the difference between request and request is within a certain range. Then, the RSU verifies the vehicle's report by examining  $e(P, \sigma_j) \overset{?}{=} e(Y_j,  H(PID_j || Y_j || C_{j1}||C_{j2}||TS))$ as $e(P,\sigma_j) = e(x_j P,H(PID_j || Y_j || C_{j1}||C_{j2}||TS)) = e(Y_j, H(PID_j || Y_j || C_{j1}||C_{j2}||TS))$. Especially, to improve efficiency, RSU can perform batch verification to check whether $e(P, \sum^{N}_{j = 1} \sigma_j) \overset{?}{=} \prod^{N}_{j=1}e(Y_j, H(PID_j || Y_j || C_{j1}||C_{j2}||TS))$, where $N$ is the number of vehicles passing by every segment. The proof is given below.

\begin{equation}
	\begin{aligned}
		e(P, \sum^{N}_{j = 1} \sigma_j)  = & e(P, \sum^N_{j=1} x_j H(PID_j || Y_j || C_{j1}||C_{j2}||TS)) \\
		& = \prod^N_{j=1} e(P, x_j  H(PID_j || Y_j || C_{j1}||C_{j2}||TS)) \\
		& = \prod^N_{j=1} e (Y_j,  H(PID_j || Y_j || C_{j1}||C_{j2}||TS)).
	\end{aligned}
\end{equation}
By performing this operation, fewer time-consuming pairing operations $e(\cdot,\cdot)$ are required (i.e., $2N \ vs. \ N+1$).

After checking the validity of vehicles' reports, the RSU executes the following steps to obtain the aggregated results in a privacy-preserving way.

\begin{itemize}
	\item {\em Step 1.} Calculate the aggregated results $C_1$ and $C_2$ based on the encrypted data $\{C_{j1}\}^{N}_{j=1}$ and $\{C_{j2}\}^{N}_{j=1}$ as follows.
	\small
	\begin{equation}
		\begin{aligned}
			\left \{ \begin{aligned}
				C_1 & =  \prod^N_{j=1} C_{j1} \ \text{mod} \ n^2   \\
				& =  \prod^N_{j=1} g^{a_1 \cdot A_1(j) + \cdots + a_M \cdot A_M(j)} \cdot r^n_{j1} \ \text{mod} \ n^2      \\
				& = g^{(a_1\sum^N_{j=1}A_1(j) + \cdots + a_M \sum^N_{j=1} A_M(j))} \cdot (\prod^{N}_{j=1}r_{j1})^n \ \text{mod} \ n^2 \\
				C_2 & = \prod^N_{j=1} C_{j2} \ \text{mod} \ n^2   \\
				& =  \prod^N_{j=1} g^{a_1\cdot S_1(j) + \cdots + a_M \cdot S_M(j)} \cdot r^n_{j2} \ \text{mod} \ n^2      \\
				& = g^{(a_1 \sum^N_{j=1} S_1(j) + \cdots + a_M \sum^N_{j=1} S_M(j))} \cdot (\prod^{N}_{j=1}r_{j2})^n \ \text{mod} \ n^2\\ \\
			\end{aligned}
			\right.
		\end{aligned}
	\end{equation}
	\small
	
	\item {\em Step 2.} Use the secret key $x_r$ to generate a 		signature as
	\begin{equation}
		\sigma_r = x_r H(ID_r||C_{1}||C_2||TS).
	\end{equation}
	
	\item {\em Step 3.} Send the aggregated and encrypted data $ID_r||C_{1}||C_2||TS||\sigma_r$ to the SP.	
\end{itemize}

\begin{figure*}
	\centering
	\includegraphics[width=10cm]{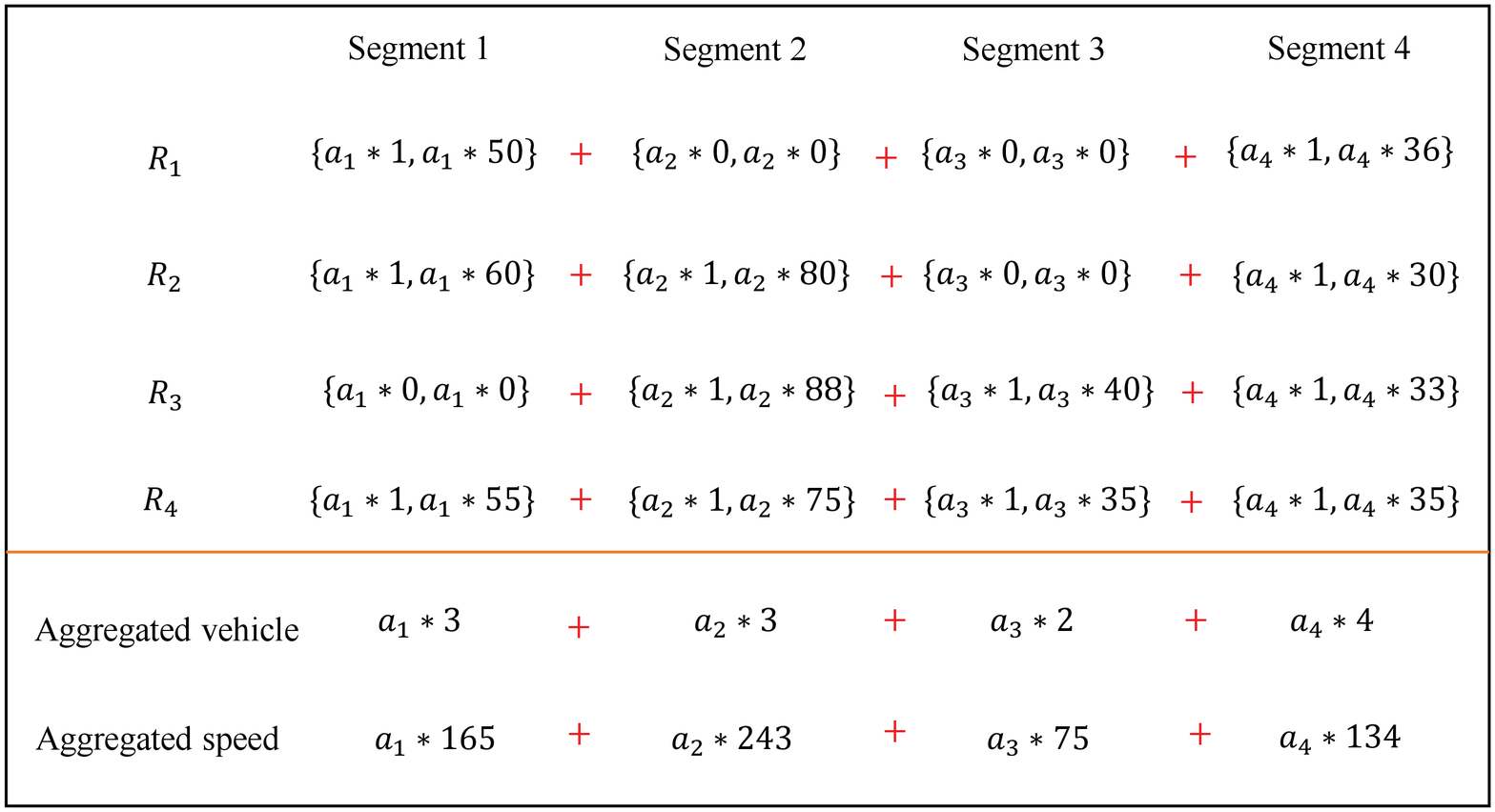}
	\caption{Traffic aggregation example in PPTM. }
	\label{fig:example}
\end{figure*}

For ease of understanding, we give an example to show how aggregated vehicle and speed are aggregated, as shown in Fig. \ref{fig:example}. The RSU receives the ciphertexts of four speed reports $\{R_1, R_2, R_3,R_4\}$, each of which contains four segments. After performing the aggregations, the aggregated results of vehicle and speed are the ciphertexts of $a_i \sum^4_{j=1} A_i(j)$ and $a_i \sum^N_{j=1} S_i(j)$ respectively, where $i \in [1,4]$. In the following, we will show how to recover the aggregated vehicle and speed for every segment.

\subsection{Secure Report Reading}
On receiving the aggregated report, SP first checks data validity by examining $e(P, \sigma_r) \overset{?}{=} e(Y_r,H(ID_r||C_{1}||C_2||TS))$, and then performs the following steps to recover the aggregated results from the ciphertexts,

\begin{itemize}
	\item {\em Step 1.} Considering $M_1 = a_1\sum^N_{j=1}A_1(j) + \cdots + a_M\sum^N_{j=1} A_M(j)$, $M_2 = a_1\sum^N_{j=1} S_1(j) + \cdots + a_M \sum^N_{j=1} S_M(j)$ and $R_1 = \prod^N_{j=1}r_{j1}, R_2 = \prod^{N}_{j=1}r_{j2}$, the aggregated ciphertexts $C_{1} = g^{M_1} \cdot R^n_1 \ \text{mod} \ n^2, C_2 = g^{M_2} \cdot R^n_2 \ \text{mod} \ n^2$ are still valid ciphertexts of Paillier cryptosystem. Hence, the SP can use the secret key $(\lambda, \mu)$ to obtain $M_1$ and $M_2$ as
	\small
	\begin{equation}
		\begin{aligned}
			\left \{ \begin{aligned}
				D(C_1) & =  a_1\sum^N_{j=1}A_1(j) + \cdots + a_M\sum^N_{j=1} A_M(j) \ \text{mod} \ n^2 \\
				D(C_2) &= a_1\sum^N_{j=1} S_1(j) + \cdots + a_M\sum^N_{j=1} S_M(j) \ \text{mod} \ n^2\\ \\
			\end{aligned}
			\right.
		\end{aligned}
	\end{equation}
	\small
	
	\item {\em Step 2.} SP then invokes algorithm 1 to recover the aggregated routes $(L_1,L_2,\cdots,L_M)$ and speed $(LS_1, LS_2, \cdots, LS_M)$, where $L_i = \sum^N_{j=1} A_i(j)$ and $LS_i = \sum^N_{j=1}S_i(j)$, $i \in [1,M]$.
\end{itemize}

\begin{algorithm}[!htbp]
	\label{A}
	\SetCommentSty{small}
	\LinesNumbered
	\caption{Recover the aggregated report}
	\KwIn{$D(C_1),D(C_2)$ and $\overrightarrow{a}$}
	\KwOut{ $\{L_i\}^{M}_{i=1}$ and $\{LS_i\}^M_{i=1}$}
	Set $\mathcal{L}_M = \sum^{M}_{i=1} \sum^N_{j=1} a_i A_i(j) \ \text{mod} \ n^2, \mathcal{LS} = \sum^M_{i=1} \sum^N_{j=1} a_i S_i(j) \ \text{mod} \ n^2$; \\
	\For{$i = M, M-1, \cdots 2$}{
		$\mathcal{L}_{i-1} = \mathcal{L}_i \ \text{mod} \ a_i$, \
		$L_i = \frac{\mathcal{L}_i - \mathcal{L}_{i-1}}{a_i};$ \
		$\mathcal{LS}_{i-1} = \mathcal{LS}_i \ \text{mod} \ a_i$, \
		$LS_i = \frac{\mathcal{LS}_i - \mathcal{LS}_{i-1}}{a_i};$ \
	}
	$L_1 = \frac{\mathcal{L}_1}{a_1};$ \
	$LS_1 = \frac{\mathcal{LS}_1}{a_1};$ \\
	\Return{$\{L_i, LS_i\}^M_{i=1}$}
\end{algorithm}

\noindent{\em The correctness of Algorithm 1.} For ease of description, we use the aggregated routes to give the correctness analysis. In this algorithm, $\mathcal{L}_M = a_1 \sum^N_{j=1} A_1(j) + a_2 \sum^N_{j=1} A_2(j) + \cdots a_{M-1} \sum^{N}_{j=1} A_{M-1}(j) + a_M \sum^{N}_{j=1}A_M(j)$. As the number of aggregated vehicle in every segment is smaller than $Q$, we have
\begin{equation}
	\begin{aligned}
		a_1 \sum^N_{j=1} A_1(j) + \cdots a_{M-1} \sum^{N}_{j=1} A_{M-1}(j)  & < (a_1 + \cdots + a_{M-1}) \cdot Q  \\
		& = \sum^{M-1}_{i=1} Q <  a_M.
	\end{aligned}
\end{equation}
Hence, $\mathcal{L}_{M-1} = \mathcal{L}_M \ \text{mod} \ a_M = a_1 \sum^{N}_{j=1}A_1(j) + \cdots + a_{M-1}\sum^{N}_{j=1} A_{M-1}(j),$ and accordingly we have
\begin{equation}
	\frac{\mathcal{L}_{M} - \mathcal{L}_{M-1}}{a_M} = \frac{a_M \sum^N_{j=1}A_M(j)}{a_M} = L_M.
\end{equation}
Following the similar analysis, $L_i = \sum^{N}_{j=1}A_i(j)$ can be proven. Also, we can prove $LS_i = \sum^{N}_{j=1} S_i(j)$, as it shares the similar procedure as $L_i$.

\subsection{Traffic Guidance and Identity Tracing}
After calculating the aggregated route and speed in all segments, i.e., $(L_1,L_2, \cdots, L_M)$ and $(LS_1, LS_2, \cdots, LS_M)$, the average speed in each segment can be computed as $L_i = \frac{LS_i}{L_i}$. At last, SP broadcasts the speed information and vehicles can select optimal routes based on the road conditions. In addition, although we assume that all vehicles report their speeds honestly, some vehicles may still upload false traffic data. In this case, the TA can periodically select some speed reports stored in the RSU and recover them to check whether they are truth or not. Since vehicles' pseudonyms are generated by using vehicles' real identity ID, malicious vehicles can be easily and quickly identified.

\section{Security Analysis}
In this section, we give the security analysis of the proposed PPTM scheme. In particular, recall the aforementioned security requirements, the analysis will focus on how our proposed PPTM scheme can protect each vehicle's report privacy, ensure report authentication and data integrity, and achieve vehicles' identity and location privacy preservation.

{\em The proposed scheme can achieve report privacy preservation.} The proposed scheme preserves reports' privacy by using the Paillier cryptosystem. In PPTM, vehicle $\mathcal{V}_j$'s location and speed are formed as $C_{j1}, C_{j2}$. Since both ciphertexts are valid ciphertexts of Paillier cryptosystem and the Paillier cryptosystem has been proven to be secure under the chosen plaintext attack, the messages are secure and privacy-preserving. That is, although an adversary may eavesdrop a ciphertext, it cannot recover the corresponding message. After receiving all reports from vehicles, instead of recovering each report, the RSU will perform report aggregation and deliver the aggregated ciphertext to the SP. Thus, even though SP holds the secret key, it can only obtain the aggregated result. Therefore, each individual vehicle's report is privacy-preserving in the proposed PPTM scheme.

{\em The proposed scheme can achieve report authentication and data integrity.} In our proposed scheme, vehicles' reports and RSU's aggregated report are signed using BLS short signature \cite{BonehLS04}. Since it has been proven that BSL short signature can defend against the CDH problem \cite{BellareR93}, our proposed scheme can guarantee the report authentication and data integrity, and any malicious behavior on the vehicles' reports will be detected.

{\em The proposed scheme can protect vehicles' identity privacy.} In our proposed scheme, vehicles periodically update their pseudonyms from TA. By changing pseudonyms, vehicles are able to keep themselves anonymous. Moreover, the proposed scheme is also effective to defend against the possible link attack presented in \cite{GolleP09}, since each vehicle's route (i.e., road segment) is aggregated and encrypted. By this way, attackers cannot infer where vehicles have been based on the given data, and accordingly cannot link their identities. Besides, although SP can obtain the aggregated route information, it is infeasible for it to recover each individual vehicle's route. Therefore, vehicles' identity privacy is preserved in the proposed PPTM scheme.

{\em The proposed scheme can protect vehicles' location privacy.} In our proposed scheme, vehicles' location privacy is preserved by aggregating their route reports. Considering the speed is location-aware, attackers may infer vehicles' locations based on the speed information. In this case, our proposed scheme is still effective, since in PPTM each individual speed is also aggregated and encrypted. Similarly, since all speed reports are also aggregated in the RSU, SP cannot obtain each individual vehicle's speed information. Thus, vehicles' location privacy is preserved.

{\em The proposed scheme can resist collusion attacks.} The basic idea to mitigate collusion attacks is to ensure the separation of data between different entities. In PPTM, with the assumption that RSU does not collude with SP, neither of them can know each individual vehicle's privacy. More specifically, the RSU cannot know vehicles' reports since they are encrypted by using the SP's public key. The SP can decrypt the summation of vehicles and speed in each segment, while not knowing each individual vehicle's data.

\section{Performance Evaluation}
In this section, we will evaluate the performance of the proposed PPTM scheme in terms of computational costs of vehicles and RSU, and communication overhead of vehicle-to-RSU and RSU-to-SP communications.

\subsection{Computational Costs}
For the proposed PPTM scheme, when a vehicle $\mathcal{V}_j$ generates an encrypted report $PID_j || Y_j || C_{j1} || C_{j2} || TS || \sigma_j$, it performs $2$ exponentiation operations in $\mathbb{Z}_{n^2}$ to calculate $C_{j1}$ and $C_{j2}$, and 1 multiplication in $\mathbb{G}$ to build the vehicle's signature $\sigma_j$. After collecting vehicles' reports, the RSU verifies the received reports with $N+1$ pairing operations. Besides, the RSU also aggregates vehicles' reports to obtain the aggregated route and speed information, which requires $N-1$ multiplication operations. However, since the multiplication operations in $\mathbb{Z}_{n^2}$ is negligible compared with the time-consuming exponentiation and pairing operations, the time costs can be omitted. In addition, to generate the signature, it also performs 1 multiplication operation in $\mathbb{G}$. As for the SP, it needs to verify the aggregated data sent from the RSU and obtain the aggregated data, which cost 1 pairing operation in $\mathbb{G}$ and 2 exponentiation operations in $\mathbb{Z}_{n^2}$ respectively. Here, we use $C_n, C_e, C_m$ to denote the computational cost of an exponentiation operation in $\mathbb{Z}_{n^2}$, a pairing operation in $\mathbb{G}$, and a multiplication operation in $\mathbb{G}$ respectively. Then, the total computation costs for the vehicle, RSU, and SP will be $2*C_n + C_m$, $(N+1)*C_e + C_m$, and $C_e + 2*C_n$ respectively.

Our proposed PPTM scheme enables each vehicle to embed its multiple speed into one compressed data, and thus large computational costs can be saved. To compare the efficiency of PPTM, a traditional approach denoted by TRPM is considered, which encrypts every individual speed information at the corresponding road segment. Under the same setting, a vehicle has to generate $M$ ciphertexts, consuming $M$ exponentiation operations in $\mathbb{Z}_{n^2}$ to perform the encryption. In addition, for the ciphertexts, the vehicle is required to generate one signature, which needs $1$ multiplication operation in $\mathbb{G}$. Thus, the total time costs will be $M*C_n + C_m$. For the RSU, it performs batch verification to authenticate the reports, which takes $N+1$ pairing operations. However, since the number of ciphertexts in TRPM is much more than that in PPTM, i.e., ($M*N$ $vs.$ $M*2$), the RSU has to perform more multiplication operations for speed aggregation. Then, the RSU generates a signature and forwards it to the RSU, which will execute $M$ exponentiation operations to recover the aggregated speed in all road segments. Thus, the total computational costs of an individual vehicle, the RSU, and the SP will be $M*C_n + C_m$, $(N+1)*C_e + C_m$, and $C_e + M*C_n$ respectively.

\begin{figure}
	\centering
	\includegraphics[width=10cm]{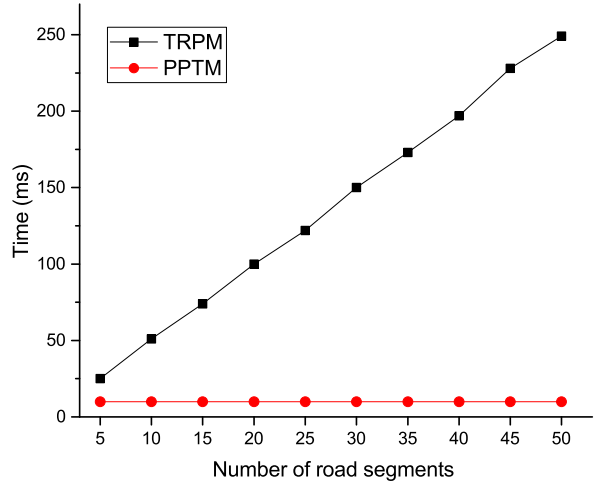}
	\caption{Computational costs of ciphertexts generation in the vehicle side. }
	\label{fig:time-vehicle}
\end{figure}

\begin{figure}
	\centering
	\includegraphics[width=10cm]{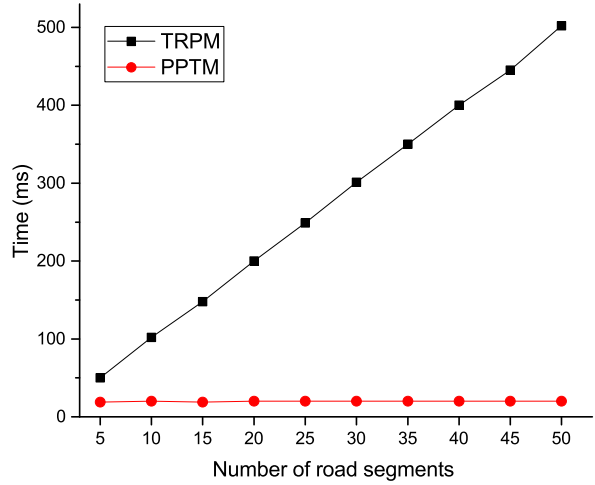}
	\caption{Computational costs of average speed calculation in the SP side. }
	\label{fig:time-SP}
\end{figure}

We list the computational costs of PPTM and TRPM in Table \ref{t1}. In addition, we conduct extensive experiments to compare the efficiency of our proposed PPTM scheme, and all experiments are performed on a laptop with Intel Core i7-7600U CPU and 16GB RAM. The security number of $\kappa$ and $\kappa_1$ are set as 1024 bits and 160 bits. The experimental results indicate that each single multiplication operation in $\mathbb{G}$ takes 2 ms, each exponentiation operation in $\mathbb{Z}_{n^2}$ takes 5 ms, and each pairing operation in $\mathbb{G}$ costs 2 ms. To validate the efficiency of our proposed PPTM, we show the computational costs in terms of road segments in Fig. \ref{fig:time-vehicle}, \ref{fig:time-SP}. From the figures, it is obvious that our proposed PPTM scheme performs much better than the traditional TRPM scheme in terms of the number of road segments, which demonstrates the correctness of the complexity analysis in Table \ref{t1}.

\begin{table}[t!]
	\caption{Comparison of computational complexity.}
	\begin{center}
		\small
		\begin{tabular}{ccccc}
			\hline
			& { PPTM} & { TRPM}    \\ \hline
			Vehicle    &   $2*C_n + C_m$  &  $M*C_n + C_m$   \\
			RSU        & $(N+1)*C_e + C_m$  &  $(N+1)*C_e + C_m$    \\
			SP         & $C_e + 2*C_n$   & $C_e + M*C_n$   \\ \hline
		\end{tabular}
	\end{center}
	\label{t1}
\end{table}

\subsection{Communication Overhead}
\begin{figure}
	\centering
	\includegraphics[width=10cm]{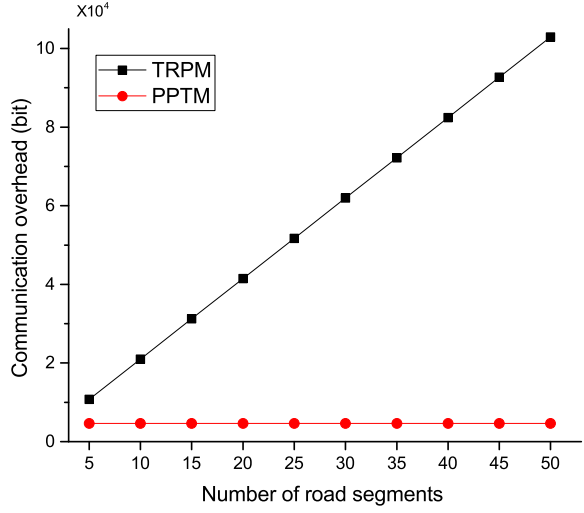}
	\caption{Communication overhead of the vehicle in the vehicle-to-RSU communication. }
	\label{fig:com-vehicle}
\end{figure}

We then analyze the communication overhead of the proposed scheme. Generally, the communications of PPTM includes two parts, i.e., vehicle-to-RSU communication and RSU-to-SP communication. For the vehicle-to-RSU communication, each individual vehicle generates its traffic report and transmits it to the RSU. Recall our previous description, the vehicle's report is defined as  $PID_j|| Y_j || C_{j1} || C_{j2} || TS || \sigma_j$ and the size is $S_v=|PID_j| + 160 + 2048*2 + |TS| + 160$, where the size of $n$ and $\mathbb{G}$ are set as 1024 bits and 160 bits respectively. RSU is responsible to collect $N$ reports in its coverage region, thus the total communication cost for the RSU is $S_R = N*S_v$. If we choose the traditional TRPM scheme, each vehicle needs to generate a ciphertext with 2048-bits for every road segment. Then, the total communication cost of vehicle-to-RSU will be $S_v=|PID_j| + 160 + 2048*M + |TS| + 160$. For comparison, we plot the bandwidth costs of both schemes in Fig. \ref{fig:com-vehicle}, where the size of $PID_j$ and $|TS|$ are both set as 100-bits. From this figure, we can see our proposed PPTM scheme can greatly save vehicles' bandwidth costs.

\begin{figure}
	\centering
	\includegraphics[width=10cm]{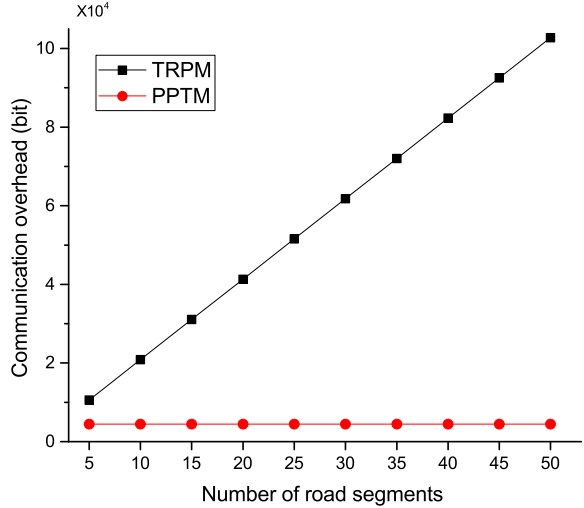}
	\caption{Communication overhead of the RSU in the RSU-to-SP communication. }
	\label{fig:com-RSU}
\end{figure}

We then consider the RSU-to-SP communication. In PPTM, RSU transmits the aggregated report $ID_r||C_{1}||C_2||TS||\sigma_r$ to the SP, which costs $S_S = |ID_r| + 2048*2 + |TS| + 160$ bits. Alternatively, TRPM needs to forward each segment's aggregated report to the SP, which requires $|ID_r| + 2048*M + |TS| + 160$ bits. Under the same setting, we compare the communication cost of both schemes in Fig. \ref{fig:com-RSU}. As can be seen, compared with the traditional TRPM scheme, our proposed PPTM scheme will aslo significantly reduce the bandwidth costs in the RSU-to-SP communication.

\section{Related Works}
Recently, traffic monitoring has received considerable attention as it is important to reduce fuel waste, air pollution, and improve drivers' driving experience. By collecting vehicles' traffic information, the traffic conditions can be better identified. Based on this, many schemes and applications have been proposed. However, the security and privacy of vehicles are still major concerns \cite{DuGXC09,XiaoRSDHG07,DuXGC07,0001DZHG07}. In fact, if drivers' privacy is not being strictly protected, they are usually reluctant to submit their data.

To realize privacy-preserving traffic monitoring, some schemes \cite{ChimYHL14,SurPR16,NiLZS16,RabiehMY17} have been proposed. For example, Chim et al. \cite{ChimYHL14} presented a secure navigation scheme which uses RSUs to guide vehicles in a distributed way. However, since all vehicles can obtain a same master key, their scheme cannot defend against the insider attacks. By using vehicular cloud and zero-knowledge proof, Sur et al. \cite{SurPR16} designed a secure navigation approach. Nevertheless, the credentials cannot be reused, which introduces more computational costs. In \cite{NiLZS16}, Ni et al. realized real-time navigation by collecting vehicles' speed information. With the technology of
randomizable signature, their scheme achieves conditional privacy preservation. Rabieh et al. \cite{RabiehMY17} further proposed a privacy-preserving route reporting scheme. In their scheme, vehicles' future routes are collected, which would be used calculate the number of vehicles appearing in next routes.

Although many efforts have been made to realize privacy-preserving traffic monitoring, most of them, nevertheless, ignore the time link attack, as described in Fig. \ref{fig:linkable}. Since vehicles are required to report their driving reports periodically or at different road segments, by linking their arriving time, vehicles' trajectories can be easily identified. That is, the traditional technologies to protect drivers' identity privacy, such as pseudonyms or randomizable signature, are not suitable in certain VANET-based applications. Inspired by the work in \cite{LuLLLS12}, we designed to use the super-increasing sequence to aggregate vehicles' routes and speed information. By this way, vehicles' identity and location privacy is preserved.

\section{Conclusion}
Vehicles' speed information is important to monitor the traffic conditions and prevent road congestion, which however threat drivers' privacy. In this paper, we propose a privacy-preserving traffic monitoring scheme by collecting vehicles' speed and route information. The main idea is to aggregate multiple speed into one compressed data so that vehicles' identity and location privacy will not be disclosed. Security analysis indicates that the proposed PPTM scheme is secure and privacy-preserving. Besides, extensive simulations demonstrate its efficiency. In the future, we will try to achieve privacy-preserving traffic monitoring without the assistance of the RSU.

\section*{References}
\bibliographystyle{plain}

\bibliography{main}

\end{document}